# *hateUS* - Analysis, impact of Social media use and Hate speech over University Student platforms: Case study, Problems, and Solutions

Naresh Kshetri, School of Business & Technology, Emporia State University, kshetrinaresh@gmail.com
Will Carter, School of Business & Technology, Emporia State University, wcarter4@g.emporia.edu
Seth Kern, School of Business & Technology, Emporia State University, skern7@g.emporia.edu
Richard Mensah, School of Business & Technology, Emporia State University, rmensah@g.emporia.edu
Bishwo Prakash Pokharel, Sault College of Applied Arts & Technology, bishwo889@hotmail.com

**Abstract** - The use of social media applications, hate speech engagement, and public debates among teenagers (primarily by university and college students) is growing day by day. The feelings of tremendous stress, anxiety, and depression via social media among our youths have a direct impact on their daily lives and personal workspace apart from delayed sleep, social media addictions, and memory loss. The use of "no phone times" and "no phone zones" is now popular in workplaces and family cultures. The use of hate speech, negotiations, and toxic words can lead to verbal abuse and cybercrime. Growing concern of mobile device security, cyberbullying, ransomware attacks, and mental health issues are another serious impact of social media among university students. The future challenges including health issues of social media use and hate speech has a serious impact on livelihood, freedom, and diverse communities of university students. Our case study is related to social media use and hate speech related to public debates over university students. We have presented the analysis and impact of social media and hate speech with several conclusions, cybercrimes, and components. The use of questionnaires for collecting primary data over university students help in the analysis of case study. The conclusion of case study and future scope of the research is extremely important to counter negative impacts.

## I. Introduction

Social media is a part of our everyday lives and has an impact on the way that the world perceives us. One of the major issues with social media today is hate speech and the effects that it can have on a person. The case study that we wanted to look at was the occurrence of hate speech on social media platforms from the perspectives of university students. Perera, S. (2023) [1] was one that used to understand how hate speech was spread over social media and why people may or may not use it to gain traction. The components and factors that were prevalent showed how hate speech is used to gain popularity. As well as the reasons that some people take part in hate speech.

Many different factors are relevant in the propagation of hate speech and the way that hate speech is spread. Sanghai S. (2023) [2] went over cyberbullying and social media addiction that led us to look at the possible correlations with hate speech and time on social media and how they may be related. Overall, the papers showed that there are many reasons for hate speech online and to understand how they are connected would give a better idea of when and where hate speech may be used. Mazari, A. C., & Kheddar, H. (2023) [3]. This paper was about setting up a data set that could capture key components that lead to hate speech. These key features were useful in understanding what factors are leading towards hate speech. From the papers listed as well as others we then went about how to get a better understanding of hate speech and impacts on university students.

We have collected and analyzed data (primary data) from a Google Form that we compiled and then asked university students to complete. We went about this by looking at other research papers online and compiling a simple set of questions that we thought would be relevant in understanding where hate speech was most prevalent the reasons for it as well as how active the university students were on social media and some other potential factors that may lead to the use of hate speech online.

The case study is organized in several sections from Section I to Section VI. Section I of the study is Introduction for social media use and hate speech over university students, along with occurrences, reasons, and problems. Section II of the study is the background study ranging from use of mobile devices, impact on mental health, link with politics, recreations, and other areas. Section III of the study is social media and hate speech as formulation for case study for this research via use of several online platforms. Section IV of the study is the case study among various university students where we collected responses from nine questions listed. Section V of the study is analysis and impact over university students showing survey findings and prevalence of hate speech in today's online world. Section VI of the study ends with conclusion and future scope regarding social media use and hate speech.

**II. Background Study**

The advent of Communication Technology tools including mobile devices (phones), computers, and the popularity of social platforms amongst young people have brought many positives in the way we live and interact with each other. For instance, young people can meet and network with new people on social platforms such as Facebook, Instagram, X (formerly, Twitter), and TikTok [4] (Mbwete, 2022). A more recent study by [4] Mbwete (2022) found that most students at all levels use social media platforms for diverse purposes including academic work, refreshment, and communication. As confirmation, [5] Dijick, 2013 elaborates on the positive influence of social media on human interactions at individual, community, and social levels. However, amongst the notable prevalent negatives of the social media age is Cyberbullying [6] (Langos, 2012). The traditional form of bullying encompasses acts including physical assault/violence, verbal, relational (social exclusion), and indirect (rumor mongering).

In relation, when these same negative acts are inflicted using computers, cell phones, and other electronic devices, then such phenomenon can be described as cyberbullying. Although [6] Langos, 2012 indicates that there is no generalized agreed-upon definition for Cyberbullying, [7] Willard 2007 defined Cyberbullying as actions that include sending or posting harmful texts, and images using the internet or other digital communication devices. Cyberbullying comes in forms such as cyberstalking, flaming, harassment, denigrating, masquerade, exclusion. Cyberbullying is considered more difficult to stop than traditional bullying because of its intensity, frequency, and unsuspecting nature [8] Patchin & Hindjuja, 2014.

For instance, studies from [9] Patchin & Hindjuja (2011), and [10] Brody & Agnew (1997) found a linkage cyberbullying often results in destructive emotions such as sadness, frustrations, anger, fear, and embarrassment all of which further leads to interpersonal violence, particularly among younger people. Furthermore, in other research works by [11] Schneider et al., 2012; [12] Ybarra, Diener-West, & Leaf, 2007, [13] Kim & Levanthal (2008), it was observed that cyberbullying is heavily associated with suicidal thoughts, low self-image, substance use, and assaultive conducts. Although cyberbullying occurs generally in cyberspace, research indicates, however, that it is prevalent among the younger generation including students.

For example, in the United Kingdom (UK), studies by [14] Oliver & Candappa (2003), found that about 6% of students between ages 12 & 13 had been bullied via nasty messages and emails; further, [15] Balding (2005) reported mobile phones have been used to bully students. In [22] Q, Li (2007) observed that about 144 out of 432 surveyed students are victims of cyberbullying. In the US, the National Center for Education Statistics (NCES) reported in 2011 that about 9% of students had indicated that they are victims of cyberbullying. Nonetheless the study by the NCES on cyberbullying focused on grades 9-12, without much emphasis on college students [16] Veronique, 2023. This has necessitated the need to investigate other students' (beyond grade 12) experiences on cyberbullying through social media platforms.

This represents an increasing concern of cyberbullying incidents among students. [17] Swearer and Hymel (2015) reveal that victims of cyberbullying are likely to experience depression, anxiety, and related health complications. Notable studies, [18] Parris et al., (2012), and [19] Konig et al., (2010) cited revenge, and jokes as some of the causes of cyberbullying among younger people. This shows that most cyberbullies often act in retaliation to the feeling of anger and frustration [20] Englander, 2008.

In addition, the anonymity of cyberspace has been found to be one of the causal factors for the rise of cyberbullying. This is because cyberbullies can hide their identities in perpetrating their actions [21] Vandabosch & Cleemput, 2008. Authors presented the cyberbullying experiences that led to cyber victimization with genders as significant roles [22]. In addition, cyberbullying is unrestrained by time and space, hence cyberbullies can attack at any time and from any place [23] Kowalski et al. 2008. Extensive research has been conducted about the consequences of cyberbullying. Against this backdrop, this study seeks to investigate college students' experiences on cyberbullying on social media platforms. This study will focus on college students primarily at Emporia State University and other universities of the United States.

**III. Social Media & Hate Speech**

Social media is one of the most prevalent forms of entertainment in the world today and connects the world on a much larger scale. It was originally developed in 1997 with SixDegrees.com and expanded and grew from there into the many different platforms and types there are today [24]. With how fast pace the world is today because of the internet many different types of social media have sprung up over time such as Facebook, twitter, reddit, YouTube, and many more. There are many different forms of social media out there however with so many options and capabilities there will always be down sides [25]. One of the most relevant and recurring components is Hate Speech.

The rise of social media with several use cases is another boost via the rise of online devices, internet population, and cybercrimes [26]. Due to the rapid technological advancements, there is an urgency to educate people about aspects of the internet along with social media use, cyber awareness as cybersecurity is a critical issue [27]. Due to that immense number of cyber-attacks increasing every year, researchers have embedded technologies like metaverse, blockchain, artificial intelligence with several legal challenges and opportunities including privacy and student experience [28] [29]. Majority of web related attacks today are due to social media use or the social engineering as the act of manipulating individuals to reveal information [30].

Hate speech is the use of abusive or threatening language to express prejudice against another person for a variety of reasons. Today, this type of speech is prevalent throughout all forms of mainstream social media as part of free speech [31]. There are a variety of different reasons that it is used so there is no one defining factor that shows that a person may or may not use hate speech. To counteract these issues many social media platforms, use community guidelines including healthcare networks for monitoring, security, and wellness [32], but that can't only do so much. So, with all these factors and information we have we compiled a set of questions aimed at trying to understand how young adults, particularly college students are affected or have seen others affected by hate speech [33]. With the intent to understand if some of these questions as ransomware attacks, cyber threats, and cryptojacking attacks are increasing [34], we can point out common components that may result in the use of hate speech.

**IV. Case Study for Social Media Use and Hate Speech**

We have presented our case study with the help of questionnaires (total nine questions) with the aim of targeting university students and their responses in terms of social media use and hate speech online. A total of 75 respondents (all university students) have answered the online questionnaire survey. For question 1, "How many social media apps have you installed on your device?" 49.3% chose 6+ social media apps [Option Three], 45.3% chose 3-6 apps [Option Two], and only 5.3% chose 1-3 apps [Option One]. For question 2 of the online survey, "Which social media site do you use frequently", 41.3% chose Facebook [Option One], 30.7% chose TikTok [Option Three], 25.3% chose Instagram [Option Two], and only 2.7% chose Other [Option Four] (please see Figure 1).

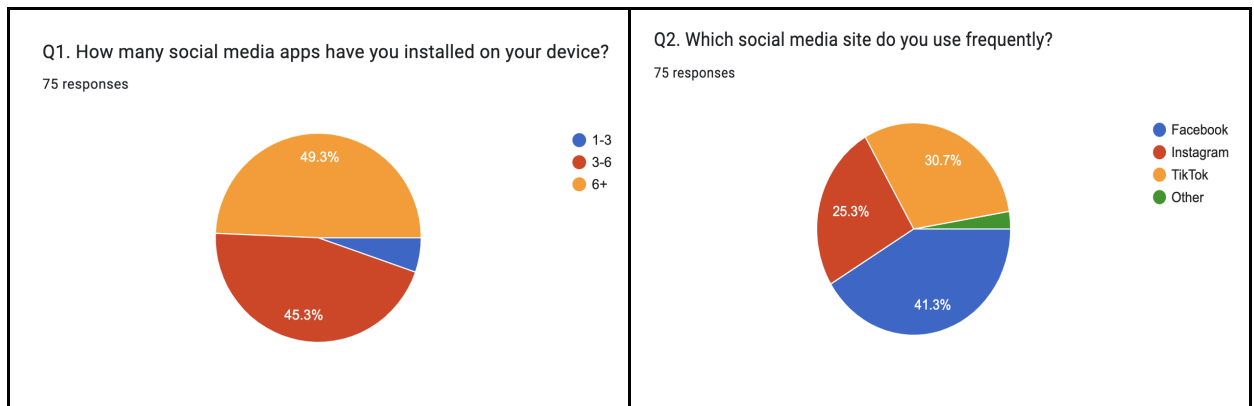

Figure 1: University student responses in pie-chart for Question 1 and Question 2 of Questionnaire

In response to question 3, "What is the major reason for your social media use?" 62.7% chose Family/Friends [Option One], 33.3% chose Expressing Self [Option Two], and only 4.0% chose Other [Option Three]. For answers to question 4, "Does social media bring you some form of Happiness?" 66.7% chose Yes [Option One], 32.0% chose Maybe [Option Three], and only 1.3% chose No [Option Two] (please see Figure 2). For question 5, "How many hours a day do you spend on social media" 64.0% chose 5-8 hours daily [Option Two], 33.3% chose 1-4 hours [Option One], and only 2.7% chose 8+ hours [Option Three].

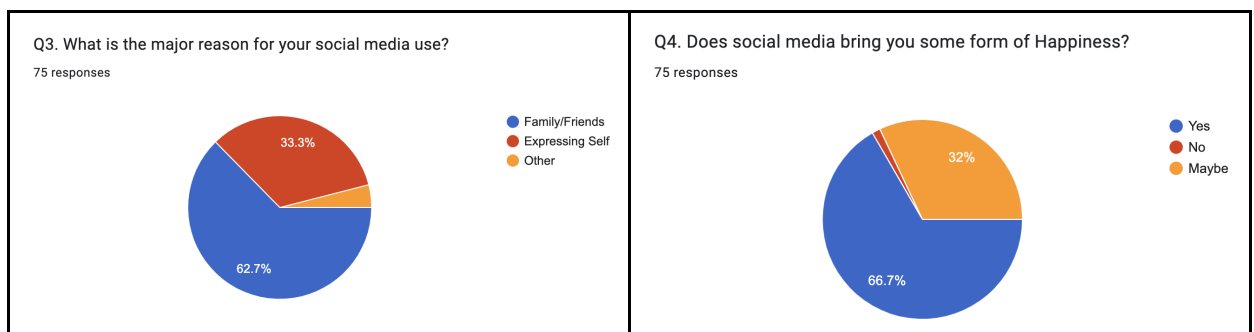

Figure 2: University student responses in pie-chart for Question 3 and Question 4 of Questionnaire

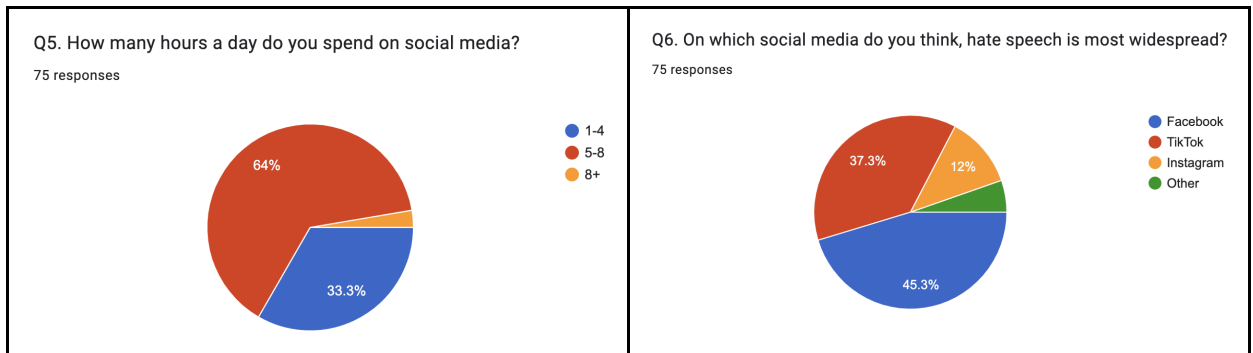

Figure 3: University student responses in pie-chart for Question 5 and Question 6 of Questionnaire

In response to question 6, "On which social media do you think, hate speech is most widespread?" 45.3% chose Facebook [Option One], 37.3% chose TikTok [Option Two], 12.0% chose Instagram [Option Three], and only 5.3% chose Other [Option Four] (please see Figure 3). For question 7, "Do you engage in public debates/discussions online?" 44.0% chose No [Option Two], 34.7% chose Maybe [Option Three], and 21.3% chose Yes [Option One]. For question 8 of the case study, "Do you believe that social media sites are doing a good job at filtering hate speech?" 65.3% chose No [Option Two], 32.0% chose Maybe [Option Three], and only 2.7% chose Yes [Option One] (please see Figure 4). For the last question of the survey, question 9, "When you see hate speech online, how often do you find it linked to politics?" 69.3% chose Very [Option One], 29.3% chose Fairly [Option Two], and only 1.3% chose Not Very [Option Three] (please see Figure 5).

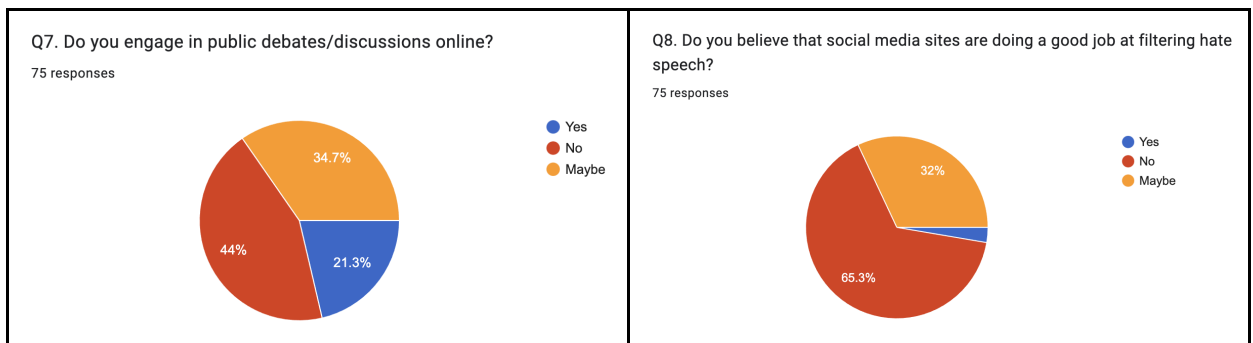

Figure 4: University student responses in pie-chart for Question 7 and Question 8 of Questionnaire

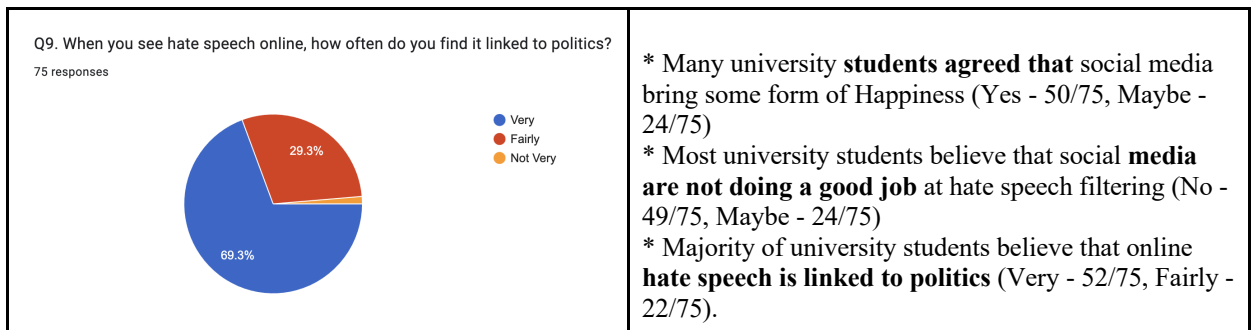

* Many university **students agreed that** social media bring some form of Happiness (Yes - 50/75, Maybe - 24/75)
* Most university students believe that social **media are not doing a good job** at hate speech filtering (No - 49/75, Maybe - 24/75)
* Majority of university students believe that online **hate speech is linked to politics** (Very - 52/75, Fairly - 22/75).

Figure 5: University student responses in pie-chart for Question 9 of Questionnaire and some findings summary

# V. Analysis and Impact over Students

Most respondents (50 out of 75) expressed that social media brings them some level of happiness. However, this engagement comes at the cost of significant time spent online, with 48 respondents reporting that they spend 5 to 8 hours per day on social media. This high usage raises concerns about potential addiction and the negative impacts associated with prolonged exposure to online environments as shown in Figure 6.

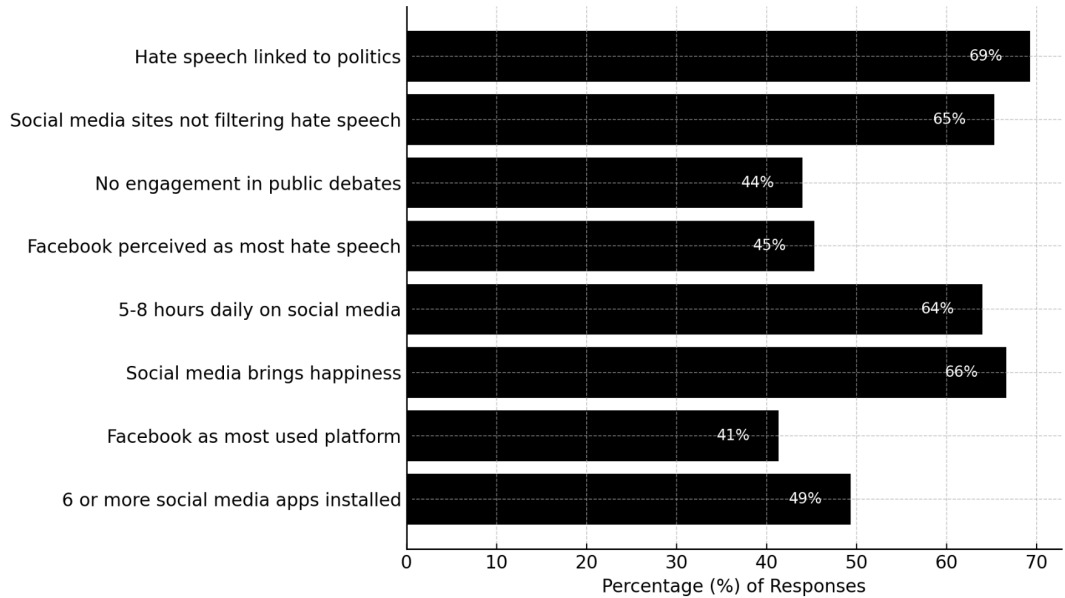

Figure 6: Survey findings from the Questionnaire via Google Forms on Social Media Use and Hate Speech among University students.

*1. Social Media Usage Among University Students*
The survey data reveals significant patterns in how university students interact with social media. Out of the 75 respondents, 37 reported having six or more social media apps on their devices, indicating that students are heavily connected across multiple platforms. Facebook emerged as the most frequently used platform, with 31 respondents identifying it as their primary site. Social media is primarily used to stay connected with family and friends, as 47 respondents indicated. This highlights the central role these platforms play in maintaining social connections among students.

*2. Prevalence of Hate Speech*
The data shows that Facebook is perceived as the platform where hate speech is mostly widespread, with 34 respondents identifying it as a hub for such behavior. Hate speech, which includes offensive, discriminatory, or harmful language directed at specific groups, is a growing concern on social platforms. Despite the large number of social media users, many students choose not to engage in public debates or discussions online, as 33 respondents indicated they avoid these interactions. This avoidance may be due to the hostile and toxic nature of many online conversations, where hate speech is prevalent.

One of the most concerning findings is that 49 respondents believe social media platforms are not doing enough to filter hate speech. This lack of confidence in the platforms' ability to regulate harmful content indicates a gap between the expectations of users and the actual effectiveness of content moderation policies. Moreover, 52 respondents noted that hate speech is often linked to politics, suggesting that political discourse is a key trigger for hate speech online.

*3. Impact on Mental Health and Well-being*
The significant time spent on social media, combined with the exposure to hate speech, has clear implications for the mental health and well-being of university students. While social media provides a sense of happiness and connection for many, it can also be a source of stress, anxiety, and depression. The presence of hate speech and toxic discussions can lead to feelings of isolation or fear of being targeted, further worsening mental health conditions among students.

Moreover, the addictive nature of social media, evidenced by the high number of hours spent online, contributes to issues like delayed sleep, memory loss, and reduced productivity. As students become more engrossed in these platforms, they may find it harder to detach from the negative content and conversations that permeate their feeds.

*4. Cyberbullying and Online Harassment*
In addition to mental health concerns, students are increasingly exposed to cyberbullying and online harassment. The prevalence of hate speech and harmful discussions creates an environment where bullying and verbal abuse can thrive. While not directly mentioned in the data, the link between hate speech and cyberbullying is well-established. Students who engage in online discussions or express their opinions publicly are often targets of harassment, which can have long-term emotional and psychological consequences [35].

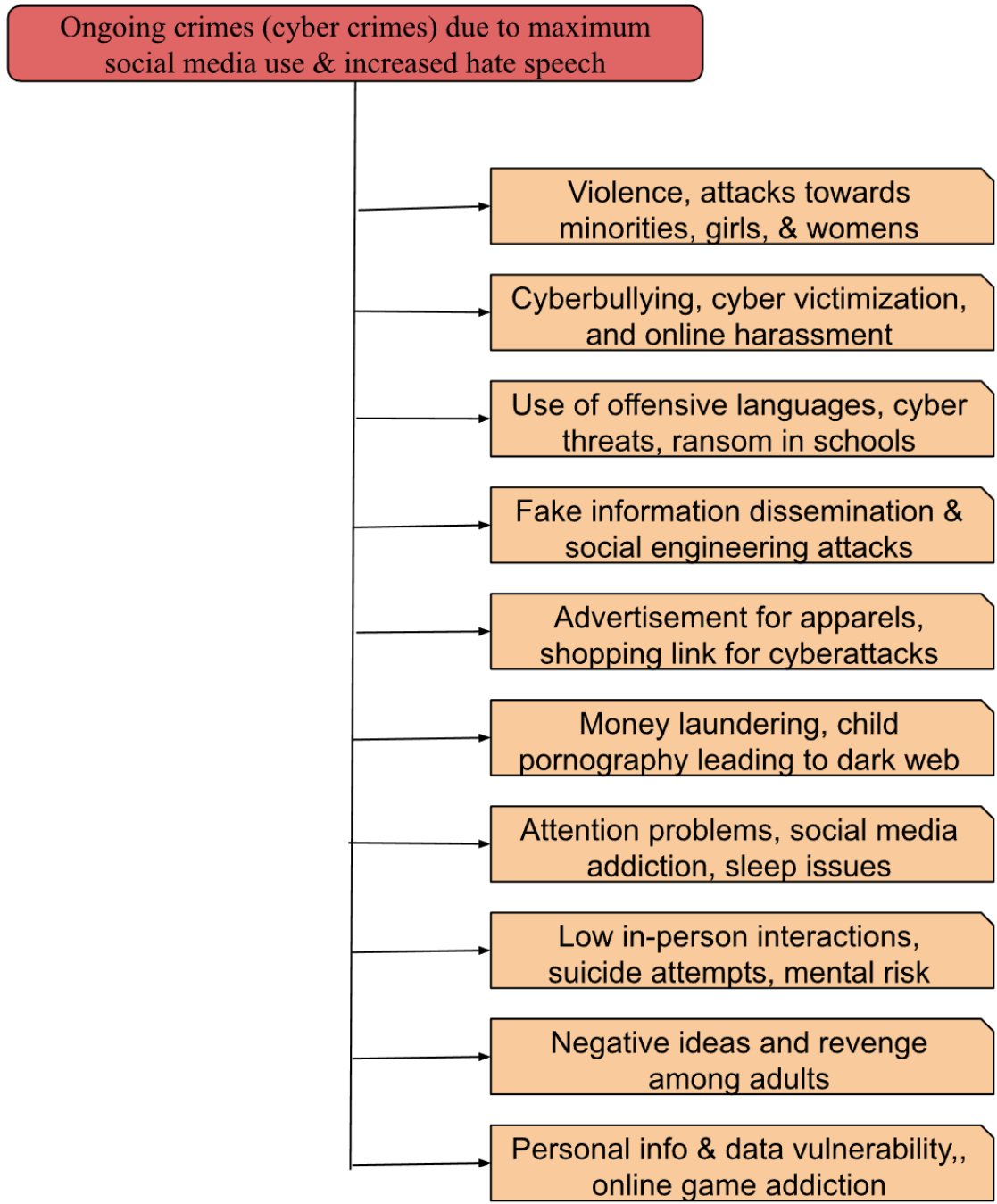

Figure 7: List of increasing online crimes including cybercrimes due to maximum Social Media use and Hate Speech among University students [2] [4] [6] [11] [15] [17] [19] [20] [25] [26] [27] [30] [31]

## VI. Conclusion & Future Scope

The case study presented over university students provided some major conclusions regarding social media use and hate speech. Some of the major conclusions are (i) Hate speech is primarily linked to politics and Facebook is the most common platform for hate speech among several social media, (ii) Social media still are not successful in filtering the hate speech content although they claim to be doing that, (iii) Students still spend 5 - 8 hours daily on social media which has serious impact on their mental health and well-being making the screen time increasing, (iv) Use of six or more social media apps installed with maximum time spend daily by students in their devices is another proof of cyberbullying and online harassment for university students.

More study regarding social media and hate speech can be conducted in future as not only related to university students. The angle of social crime and individual depression along with concerns of cyberbullying among several citizens can be analyzed. The future scope of this study can also be upon working professionals regarding device usage and cyber-attacks not limited to phishing attacks and ransomware attacks to employees.